\documentclass[prd,superscriptaddress,twocolumn,showpacs,preprintnumbers,amsmath,amssymb,altaffilletter]{revtex4}

\usepackage{graphicx}
\usepackage{dcolumn}
\usepackage{bm}

\makeatletter
\def\@fnsymbol#1{\ifcase#1\or * \or  $+$ \or  \$ \or \#  \or \dag \or \ddag \or
$\mathsection$ \or $ \mathparagraph$ \or $\|$  \or
\textordfeminine \or \textbullet   \or ** \or $++$ \or  \$\$ \or
\#\#  \or \dag\dag \or \ddag\ddag \or $\mathsection\mathsection$
\or $ \mathparagraph\mathparagraph$ \or $\|\|$  \or
\textordfeminine\textordfeminine \or \textbullet \textbullet \or
*** \or $+++$ \or  \$\$\$ \or \#\#  \or \dag\dag \or \ddag\ddag
\or $\mathsection \mathsection\mathsection$ \or $ \mathparagraph
\mathparagraph\mathparagraph$ \or $\|\|\|$  \or
\textordfeminine\textordfeminine\textordfeminine \or
\textbullet\textbullet\textbullet \or \else \@ctrerr\fi}
\makeatother

\newcommand{\xvec}[0]{{\stackrel{\rightharpoonup}{x}}}

\newcommand{\dxvec}[0]{{\Delta \!\! \stackrel{\rightharpoonup}{x}}}
\newcommand{\omg}[0]{{\hat{\Omega}}}

\def\lsim{\mathrel{\rlap{\lower4pt\hbox{\hskip1pt$\sim$}}
    \raise1pt\hbox{$<$}}}                
\def\gsim{\mathrel{\rlap{\lower4pt\hbox{\hskip1pt$\sim$}}
    \raise1pt\hbox{$>$}}}                

\newcommand{\e}[1]{\ensuremath{\times 10^{#1}}}
\newcommand{\hzff}[1]{\ensuremath{{\rm Hz}^{-1}\left(\frac{f}{100{\rm Hz}}\right)^{#1}}}
\newcommand{\ff}[1]{\ensuremath{\left(\frac{f}{100{\rm Hz}}\right)^{#1}}}
\newcommand{\hz}[0]{\ensuremath{~{\rm Hz}}}

\newcommand{\h}[0]{\ensuremath{{\rm h}}}
\newcommand{\m}[0]{\ensuremath{{\rm m}}}
\newcommand{\s}[0]{\ensuremath{{\rm s}}}

\begin{document}
\preprint{LIGO-P060029-00-Z}

\title{Upper limit map of a background of gravitational waves}

\newcommand*{\AG}{Albert-Einstein-Institut, Max-Planck-Institut f\"ur Gravitationsphysik, D-14476 Golm, Germany}                                      \affiliation{\AG}                                 
\newcommand*{\AH}{Albert-Einstein-Institut, Max-Planck-Institut f\"ur Gravitationsphysik, D-30167 Hannover, Germany}                                  \affiliation{\AH}                                 
\newcommand*{\AU}{Andrews University, Berrien Springs, MI 49104 USA}                                                                                  \affiliation{\AU}                                 
\newcommand*{\AN}{Australian National University, Canberra, 0200, Australia}                                                                          \affiliation{\AN}                                 
\newcommand*{\CH}{California Institute of Technology, Pasadena, CA  91125, USA}                                                                       \affiliation{\CH}                                 
\newcommand*{\CA}{Caltech-CaRT, Pasadena, CA  91125, USA}                                                                                             \affiliation{\CA}                                 
\newcommand*{\CU}{Cardiff University, Cardiff, CF2 3YB, United Kingdom}                                                                               \affiliation{\CU}                                 
\newcommand*{\CL}{Carleton College, Northfield, MN  55057, USA}                                                                                       \affiliation{\CL}                                 
\newcommand*{\CS}{Charles Sturt University, Wagga Wagga, NSW 2678, Australia}                                                                         \affiliation{\CS}                                 
\newcommand*{\CO}{Columbia University, New York, NY  10027, USA}                                                                                      \affiliation{\CO}                                 
\newcommand*{\ER}{Embry-Riddle Aeronautical University, Prescott, AZ   86301 USA}                                                                     \affiliation{\ER}                                 
\newcommand*{\HC}{Hobart and William Smith Colleges, Geneva, NY  14456, USA}                                                                          \affiliation{\HC}                                 
\newcommand*{\IU}{Inter-University Centre for Astronomy  and Astrophysics, Pune - 411007, India}                                                      \affiliation{\IU}                                 
\newcommand*{\CT}{LIGO - California Institute of Technology, Pasadena, CA  91125, USA}                                                                \affiliation{\CT}                                 
\newcommand*{\LM}{LIGO - Massachusetts Institute of Technology, Cambridge, MA 02139, USA}                                                             \affiliation{\LM}                                 
\newcommand*{\LO}{LIGO Hanford Observatory, Richland, WA  99352, USA}                                                                                 \affiliation{\LO}                                 
\newcommand*{\LV}{LIGO Livingston Observatory, Livingston, LA  70754, USA}                                                                            \affiliation{\LV}                                 
\newcommand*{\LU}{Louisiana State University, Baton Rouge, LA  70803, USA}                                                                            \affiliation{\LU}                                 
\newcommand*{\LE}{Louisiana Tech University, Ruston, LA  71272, USA}                                                                                  \affiliation{\LE}                                 
\newcommand*{\LL}{Loyola University, New Orleans, LA 70118, USA}                                                                                      \affiliation{\LL}                                 
\newcommand*{\MS}{Moscow State University, Moscow, 119992, Russia}                                                                                    \affiliation{\MS}                                 
\newcommand*{\ND}{NASA/Goddard Space Flight Center, Greenbelt, MD  20771, USA}                                                                        \affiliation{\ND}                                 
\newcommand*{\NA}{National Astronomical Observatory of Japan, Tokyo  181-8588, Japan}                                                                 \affiliation{\NA}                                 
\newcommand*{\NO}{Northwestern University, Evanston, IL  60208, USA}                                                                                  \affiliation{\NO}                                 
\newcommand*{\RI}{Rochester Institute of Technology, Rochester, NY 14623, USA}                                                                        \affiliation{\RI}                                 
\newcommand*{\RA}{Rutherford Appleton Laboratory, Chilton, Didcot, Oxon OX11 0QX United Kingdom}                                                      \affiliation{\RA}                                 
\newcommand*{\SJ}{San Jose State University, San Jose, CA 95192, USA}                                                                                 \affiliation{\SJ}                                 
\newcommand*{\SE}{Southeastern Louisiana University, Hammond, LA  70402, USA}                                                                         \affiliation{\SE}                                 
\newcommand*{\SO}{Southern University and A\&M College, Baton Rouge, LA  70813, USA}                                                                  \affiliation{\SO}                                 
\newcommand*{\SA}{Stanford University, Stanford, CA  94305, USA}                                                                                      \affiliation{\SA}                                 
\newcommand*{\SR}{Syracuse University, Syracuse, NY  13244, USA}                                                                                      \affiliation{\SR}                                 
\newcommand*{\PU}{The Pennsylvania State University, University Park, PA  16802, USA}                                                                 \affiliation{\PU}                                 
\newcommand*{\TC}{The University of Texas at Brownsville and Texas Southmost College, Brownsville, TX  78520, USA}                                    \affiliation{\TC}                                 
\newcommand*{\TR}{Trinity University, San Antonio, TX  78212, USA}                                                                                    \affiliation{\TR}                                 
\newcommand*{\HU}{Universit{\"a}t Hannover, D-30167 Hannover, Germany}                                                                                \affiliation{\HU}                                 
\newcommand*{\BB}{Universitat de les Illes Balears, E-07122 Palma de Mallorca, Spain}                                                                 \affiliation{\BB}                                 
\newcommand*{\UA}{University of Adelaide, Adelaide, SA 5005, Australia}                                                                               \affiliation{\UA}                                 
\newcommand*{\BR}{University of Birmingham, Birmingham, B15 2TT, United Kingdom}                                                                      \affiliation{\BR}                                 
\newcommand*{\FA}{University of Florida, Gainesville, FL  32611, USA}                                                                                 \affiliation{\FA}                                 
\newcommand*{\GU}{University of Glasgow, Glasgow, G12 8QQ, United Kingdom}                                                                            \affiliation{\GU}                                 
\newcommand*{\MD}{University of Maryland, College Park, MD 20742 USA}                                                                                 \affiliation{\MD}                                 
\newcommand*{\MU}{University of Michigan, Ann Arbor, MI  48109, USA}                                                                                  \affiliation{\MU}                                 
\newcommand*{\OU}{University of Oregon, Eugene, OR  97403, USA}                                                                                       \affiliation{\OU}                                 
\newcommand*{\RO}{University of Rochester, Rochester, NY  14627, USA}                                                                                 \affiliation{\RO}                                 
\newcommand*{\SL}{University of Salerno, 84084 Fisciano (Salerno), Italy}                                                                             \affiliation{\SL}                                 
\newcommand*{\SN}{University of Sannio at Benevento, I-82100 Benevento, Italy}                                                                        \affiliation{\SN}                                 
\newcommand*{\US}{University of Southampton, Southampton, SO17 1BJ, United Kingdom}                                                                   \affiliation{\US}                                 
\newcommand*{\SC}{University of Strathclyde, Glasgow, G1 1XQ, United Kingdom}                                                                         \affiliation{\SC}                                 
\newcommand*{\WS}{University of Washington, Seattle, WA, 98195}                                                                                       \affiliation{\WS}                                 
\newcommand*{\WA}{University of Western Australia, Crawley, WA 6009, Australia}                                                                       \affiliation{\WA}                                 
\newcommand*{\UW}{University of Wisconsin-Milwaukee, Milwaukee, WI  53201, USA}                                                                       \affiliation{\UW}                                 
\newcommand*{\WU}{Washington State University, Pullman, WA 99164, USA}                                                                                \affiliation{\WU}                                 
\author{B.~Abbott}    \affiliation{\CT}                                                                                                                                                                 
\author{R.~Abbott}    \affiliation{\CT}                                                                                                                                                                 
\author{R.~Adhikari}    \affiliation{\CT}                                                                                                                                                               
\author{J.~Agresti}    \affiliation{\CT}                                                                                                                                                                
\author{P.~Ajith}    \affiliation{\AH}                                                                                                                                                                  
\author{B.~Allen}    \affiliation{\AH}  \affiliation{\UW}                                                                                                                                               
\author{R.~Amin}    \affiliation{\LU}                                                                                                                                                                   
\author{S.~B.~Anderson}    \affiliation{\CT}                                                                                                                                                            
\author{W.~G.~Anderson}    \affiliation{\UW}                                                                                                                                                            
\author{M.~Arain}    \affiliation{\FA}                                                                                                                                                                  
% ...

\collaboration{The LIGO Scientific Collaboration, http://www.ligo.org}
\noaffiliation

\date{\today}

\begin{abstract}
We searched for an anisotropic
background of gravitational waves using data from the LIGO S4 science run
and a method that is optimized for point sources.
This is appropriate if, for example, the
gravitational wave background is dominated by a small number of distinct astrophysical sources.
No signal was seen. Upper limit maps were produced
assuming two different power laws for
the source strain power spectrum. For an $f^{-3}$ power law and using
the $50{\hz}$ to $1.8{\rm kHz}$ band
the upper limits on the source strain power spectrum vary between
$1.2\e{-48} {\rm Hz}^{-1} \left({100~{\rm Hz}}/{f}\right)^3$
and
$1.2\e{-47} {\rm Hz}^{-1} \left({100~{\rm Hz}}/{f}\right)^3$,
depending on the position in the sky.
Similarly, in the case of constant strain power spectrum, the upper limits vary between
$8.5\e{-49} {\rm Hz}^{-1}$
and
$6.1\e{-48} {\rm Hz}^{-1}$.
As a side product a limit on an isotropic background of gravitational waves
was also obtained.
All limits are at the 90\% confidence level.
Finally, as an application, we focused on the direction of Sco-X1, the closest low-mass
X-ray binary.
We compare the upper limit on strain amplitude obtained by this method to
expectations based on the X-ray luminosity of Sco-X1. 

\end{abstract}

\maketitle

\section{\label{sec:Introduction}Introduction}

A stochastic background of gravitational waves can be non-isotropic
if, for example,
the dominant source of stochastic gravitational waves
comes from an ensemble of astrophysical sources (e.g. \cite{tania,bildsten}), and
if this ensemble is dominated by its strongest members.
So far the LIGO Scientific Collaboration has analyzed the data from the first science runs
for a stochastic background of gravitational waves \cite{vuk,s3stoch,s1stoch}, assuming that this
background is {\it isotropic}.
If astrophysical sources indeed dominate
this background, one should look for anisotropies. 

A method that is optimized for extreme anisotropies, namely point sources of stochastic
gravitational radiation,
was presented in \cite{ballmer}. It is based on the cross-correlation
of the data streams from two spatially separated gravitational wave interferometers,
and is referred to as radiometer analysis.
We have analyzed the data of the 4th LIGO science run using this method.

Section \ref{sec:Radiometer} is a short description of the radiometer analysis method.
The peculiarities of the S4 science run are summarized in section \ref{sec:S4},
and we discuss the results in section \ref{sec:Results}.

\section{\label{sec:Radiometer}Method description}

A stochastic background of gravitational waves can be distinguished from other sources of detector
noise by cross-correlating two independent detectors.
Thus we cross-correlate the data streams from a pair of detectors
with a
cross-correlation kernel $Q$,
chosen to be optimal for
a source which is specified by
an assumed strain power spectrum
$H(f)$ and angular power distribution $P(\omg)$.
Specifically, with $\tilde{s}_1(f)$ and
$\tilde{s}_2(f)$ representing the Fourier transforms of the strain
outputs of two detectors, this cross-correlation is computed in the frequency domain segment by segment as:
\begin{equation}\label{e:xcorr}
    Y_t = \int_{-\infty}^{\infty}df
        \,\tilde{s}_1^*(f)\,Q_t(f)\,\tilde{s}_2(f).
\end{equation}
In contrast to the isotropic analysis the optimal filter
$Q_t$ is now sidereal time
dependent. It has the
general form:
\begin{equation}\label{e:optfilt}
Q_t(f) = \lambda_t \frac{\int_{S^2} d\omg \gamma_{\omg,t}(f) P(\omg) H(f)}{P_1(f) P_2(f)}
\end{equation}
where $\lambda_t$ is a normalization factor, $P_1$ and $P_2$ are the
strain noise power spectra of the two detectors, $H$ is
the strain power spectrum of the stochastic signal we
search for, and the factor
$\gamma_{\omg,t}$ takes into 
account 
the sidereal time dependent time delay due to the detector separation and
the directionality of the acceptance of the detector pair.
Assuming that the source is unpolarized, $\gamma_{\omg,t}$ is given by:
\begin{equation}\label{e:gammaOmegaRM}
\gamma_{\omg,t}(f)=\frac{1}{2} \sum_A e^{i 2 \pi f \omg \cdot \frac{\dxvec}{c}}
    \,\, F_1^A(\omg) F_2^A(\omg)
\end{equation}
where $\dxvec=\xvec_2 - \xvec_1$ is the detector separation vector,
$\omg$ is the unit vector specifying the sky position and 
\begin{equation}\label{e:F}
    F_i^A(\omg) = e_{ab}^A(\omg) \frac{1}{2}(\hat{X}_i^a \hat{X}_i^b - \hat{Y}_i^a \hat{Y}_i^b)
\end{equation}
is the response of detector $i$ to a zero frequency, unit amplitude, $A=+\,{\rm or}\,\times$ polarized gravitational wave.
$e_{ab}^A(\omg)$ is the spin-two polarization tensor for polarization $A$
and $\hat{X}_i^a$ and $\hat{Y}_i^a$ are unit vectors pointing in the directions
of the detector arms (see \cite{allenromano} for details).
The sidereal time dependence enters through the rotation of the earth,
affecting $\hat{X}_i^a$, $\hat{Y}_i^a$ and  $\dxvec$.

The optimal filter $Q_t$ is derived assuming that the intrinsic detector
noise is Gaussian and stationary over the measurement time,
uncorrelated between detectors, and uncorrelated with and much
greater in power than the stochastic gravitational wave signal. Under these
assumptions the expected variance, $\sigma^2_{Y_t}$, of the
cross-correlation is dominated by the noise in the individual
detectors, whereas the expected value of the cross-correlation $Y_t$
depends on the stochastic background power spectrum:
\begin{equation}\label{e:sigma}
    \sigma_{Y_t}^2 \equiv \langle Y_t^2 \rangle - \langle Y_t \rangle^2 \approx
               \frac{T}{4} \left( Q_t,Q_t \right)
\end{equation}
\begin{equation}\label{e:mu}
    \langle Y_t \rangle = T \left( Q_t,
    \frac{\int_{S^2} d\omg \gamma_{\omg,t} P(\omg) H}{P_1 P_2}
    \right)
\end{equation}
Here the scalar product $\left(\cdot,\cdot\right)$ is defined as
$\left(A,B\right) = \int_{-\infty}^{\infty} A^*(f) B(f) P_1(f) P_2(f) df$
and $T$ is the duration of the measurement.

Equation \ref{e:optfilt} defines the optimal filter $Q_t$
for any arbitrary choice of $P(\omg)$. 
To optimize the method for finite, but unresolved
astrophysical sources
one should use a $P(\omg)$ that covers only a localized patch in the sky.
But the angular resolution is diffraction limited with the detector
separation as baseline and the frequency content weighted by $H^2 P_1^{-1} P_2^{-1}$.
For a constant $H(f)$ this corresponds to a resolution of
several tens of square degrees,
so that astrophysical sources will not be spatially resolved. Thus we chose
to optimize the method for true point
sources, i.e. $P(\omg)=\delta^2(\omg,\omg')$, which also allows for a
more efficient implementation (see \cite{ballmer}).

We define the strain power spectrum $H(f)$ of a point source
as one-sided (positive frequencies only)
and including the power in both polarizations. Thus $H(f)$ is
related to the gravitational luminosity $L_{GW}$ and the gravitational energy flux $F(f)$ through
\begin{equation}
\label{eq:StraintoLuminosity}
L_{GW} = \int_{f_{\rm min}}^{f_{\rm max}} F(f) df = \frac{c^3 \pi}{4 G} \int_{f_{\rm min}}^{f_{\rm max}} H(f) f^2 df,
\end{equation}
with $c$ the light speed and $G$ Newton's constant.
We look for strain power spectra $H(f)$ in the form of a power law with exponent $\beta$.
The amplitude at the pivot point of $100\hz$ is described by $H_{\beta}$, i.e.
\begin{equation}\label{e:Hbeta}
H(f)=H_{\beta} \left( \frac{f}{100\hz} \right)^{\beta}.
\end{equation}
With this definition we can choose the normalization of the optimal filter $Q_t$ such
that equation \ref{e:mu} reduces to
\begin{equation}\label{e:mu2}
    \langle Y_t \rangle = H_{\beta}.
\end{equation}

The data set from a
given interferometer pair is divided into equal-length intervals,
and the cross-correlation $Y_t$ and theoretical $\sigma_{Y_t}$ are
calculated for each interval, yielding a set $\{Y_t,
\sigma_{Y_t}\}$ of such values for each sky direction $\omg$, with $t$ the mid-segment sidereal time.
The optimal filter $Q_t$ is kept constant and equal to its mid-segment value
for the whole segment.
The remaining error due to this discretization is of second order in
$(T_{\rm seg}/{1~{\rm day}})$ and is given by:
\begin{equation}\label{e:errorRM}
\begin{split}
Y_{\rm err}(T_{\rm seg})/Y &=
\frac{{T_{\rm seg}}^2}{24} \frac{
\int_{-\infty}^{\infty} \frac{\partial^2\gamma_{\omg'}^*}{\partial t^2} \gamma_{\omg'}
\frac{H^2}{P_1 P_2} df
}{
\int_{-\infty}^{\infty} \left| \gamma_{\omg'} \right|^2 \frac{H^2}{P_1 P_2} df
} \\&= 
O \left( \left( \frac{2 \pi f d}{c} \frac{T_{\rm seg}}{1~{\rm day}} \right)^2 \right)
\end{split}
\end{equation}
with $f$ the typical frequency and $d$ the detector separation.
At the same time the interval length can be
chosen such that the detector noise is relatively
stationary over one interval.
We use an interval length of 60~sec, which guarantees that 
the relative error $Y_{\rm err}(T_{\rm seg})/Y$ is less than 1\%.
The cross-correlation values
are combined to produce a final cross-correlation estimator,
$Y_{\rm opt}$, that maximizes the signal-to-noise ratio, and has
variance $\sigma_{\rm opt}^2$:
\begin{equation}\label{e:Yopt}
  \begin{array}{cc}
    Y_{\rm opt} = \sum_t \sigma_{Y_t}^{-2} Y_t / \sigma_{\rm
    opt}^{-2}\ ,
    &
\,\,\,\,\,\,\,\,\,\,\,\,
    \sigma_{\rm opt}^{-2} = \sum_t \sigma_{Y_t}^{-2}\ .
  \end{array}
\end{equation}
In practice the intervals are overlapping by 50\% to avoid
the effective loss of half the data due to the required windowing (Hanning). Thus
equation \ref{e:Yopt} was modified slightly to take the correlation of
neighboring segments into account.

\begin{figure}[htbp!]
\begin{center}
  \includegraphics[width=1\linewidth]{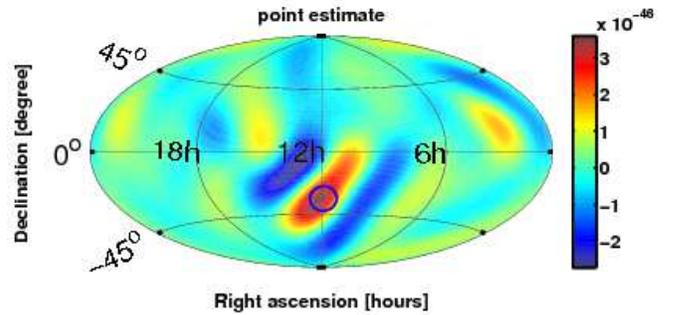}
  \caption[Injected pulsar \#3]{{\bf Injected pulsar \#3:}
           The analysis was run using the $108.625\hz - 109.125\hz$ frequency band.
	   The artificial signal of Pulsar \#3 at $108.86\hz$ stands out with a signal-to-noise ratio of $9.2$.
           The circle marks the position of the simulated pulsar.}
  \label{fig:HWIPulsar3}
\end{center}
\end{figure}

The data was downsampled to $4096\hz$ and high-pass
filtered with a sixth order Butterworth filter with a cut-off frequency at $40\hz$.
Frequencies between $50\hz$ and $1800\hz$ were used for the analysis and the frequency
bin width was $0.25\hz$. Frequency bins around multiples of $60\hz$ up to the tenth harmonic
were removed, along with bins
near a set of nearly monochromatic injected signals used to 
simulate pulsars. These artificial pulsars proved useful
in a separate end-to-end check of this analysis pipeline, which
successfully recovered the sky locations, frequencies and 
strengths of three such pulsars listed in TABLE \ref{t:injpulsar}.
The resulting map for one of these pulsars is shown in FIG. \ref{fig:HWIPulsar3}.

\begin{table*}
\begin{ruledtabular}
\begin{tabular}{llll}
\multicolumn{4}{l}
{{\bf Injected pulsars}}\\ 
Quantity & Pulsar \#3 & Pulsar \#4 & Pulsar \#8 \\ \hline 
Frequency during S4 run &$108.86~{\rm Hz}$&$1402.20~{\rm Hz}$&$193.94~{\rm Hz}$\\
Noise level ($\sigma$) &$1.89\e{-47}$    &$6.04\e{-46}$     &$1.73\e{-47}$	\\
Injected $H df$ (corrected for polarization) &$1.74\e{-46}$   &$4.28\e{-44}$	&$1.54\e{-46}$\\
Recovered $H df$ on source &$1.74\e{-46}$   &$4.05\e{-44}$        &$1.79\e{-46}$\\
Signal-to-noise ratio (SNR) &$9.2$    	  &$67.1$	     &$10.3$\\
Injected position &$11\h~53\m~29.4\s$  &$18\h~39\m~57.0\s$   &$23\h~25\m~33.5\s$	\\
                  &$-33^{\circ}~26'~11.8''$&$-12^{\circ}~27'~59.8''$ &$-33^{\circ}~25'~6.7''$\\
Recovered position (max SNR)    &$12\h~12\m$        &$18\h~40\m$         &$23\h~16\m$ 	\\
                   		&$-37^{\circ}$           &$-13^{\circ}$            &$-32^{\circ}$ 		\\
\end{tabular}
\end{ruledtabular}
\caption[Injected pulsars]{
          {\bf Injected pulsars:} The table shows the level at which the three
	  strongest injected pulsars were recovered. $H df$ denotes the RMS strain power
	  over the $0.5\hz$ band that was used. The reported values of for the injected $H df$
	  include corrections that account for the difference between the polarized pulsar injection
	  and an unpolarized source that is expected by the analysis. The underestimate of Pulsar
	  \#4 is due to a known bias of the analysis method in the case of a signal strong enough to
	  affect the power spectrum estimation.}
\label{t:injpulsar}
\end{table*}

\section{\label{sec:S4}The LIGO S4 science run}

The LIGO S4 science run consisted of one month of coincident data taking with all three
LIGO interferometers (22 Feb 2005 noon to 23 Mar 2005 midnight CST). During that time all three interferometers
where roughly a factor of 2 in amplitude
away from design sensitivity over almost the whole frequency band.
Also, the Livingston interferometer was equipped with a Hydraulic External Pre-Isolation
(HEPI) system, allowing it to stay locked during day time. This made S4 the first LIGO
science run with all-day coverage at both sites. A more detailed description of the
LIGO interferometers is given in \cite{s1ligo}.

Since the radiometer analysis requires two spatially separated sites we used only data
from the two 4 km interferometers (H1 in Hanford and L1 in Livingston).
For these two interferometers 
about 20 days of coincident data was collected, corresponding to a duty factor of 69\%.

\begin{figure}[!thb]
\centering
\begin{minipage}[t]{2.75in}
\includegraphics[width=1\linewidth]{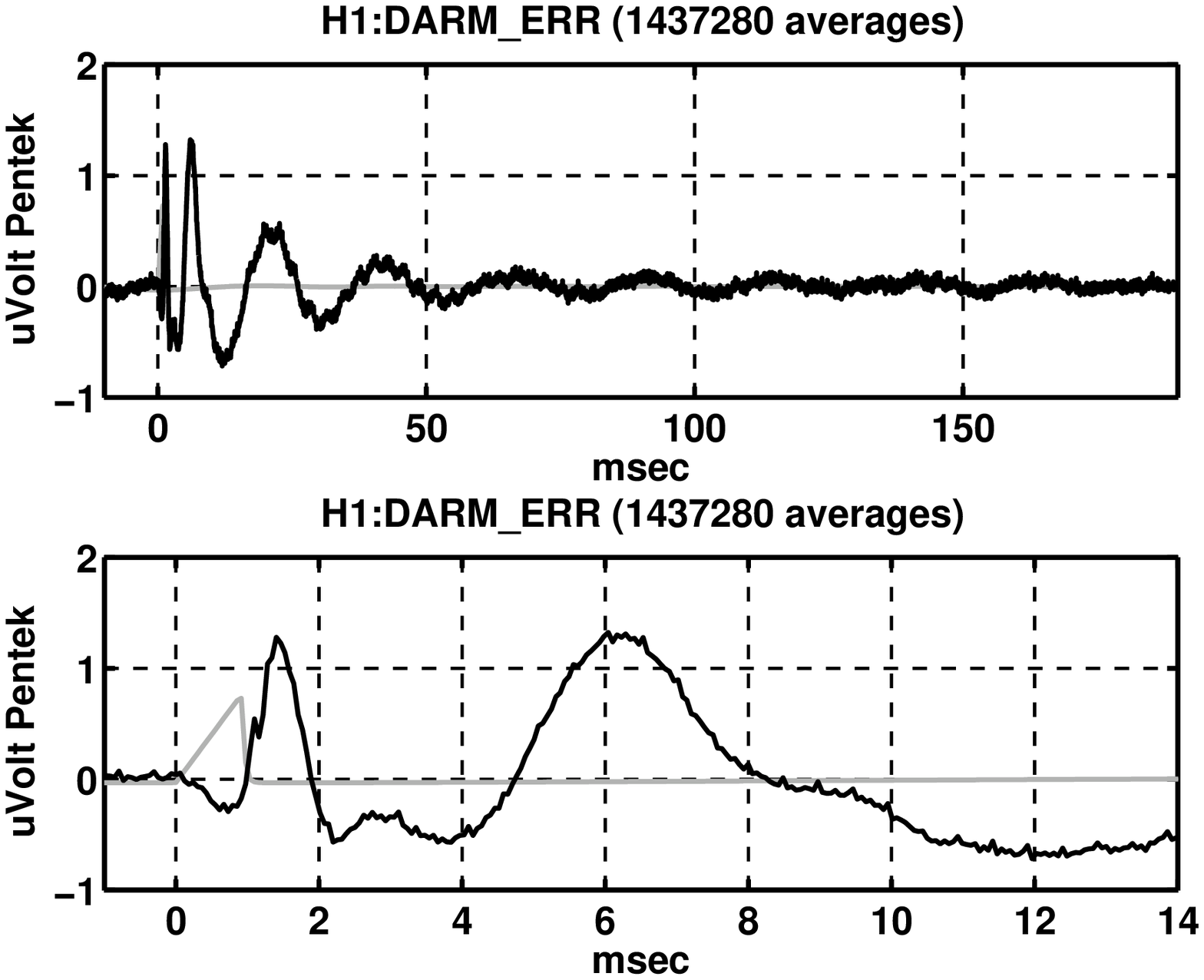}
\end{minipage}
\hfill
\begin{minipage}[t]{2.75in}
\includegraphics[width=1\linewidth]{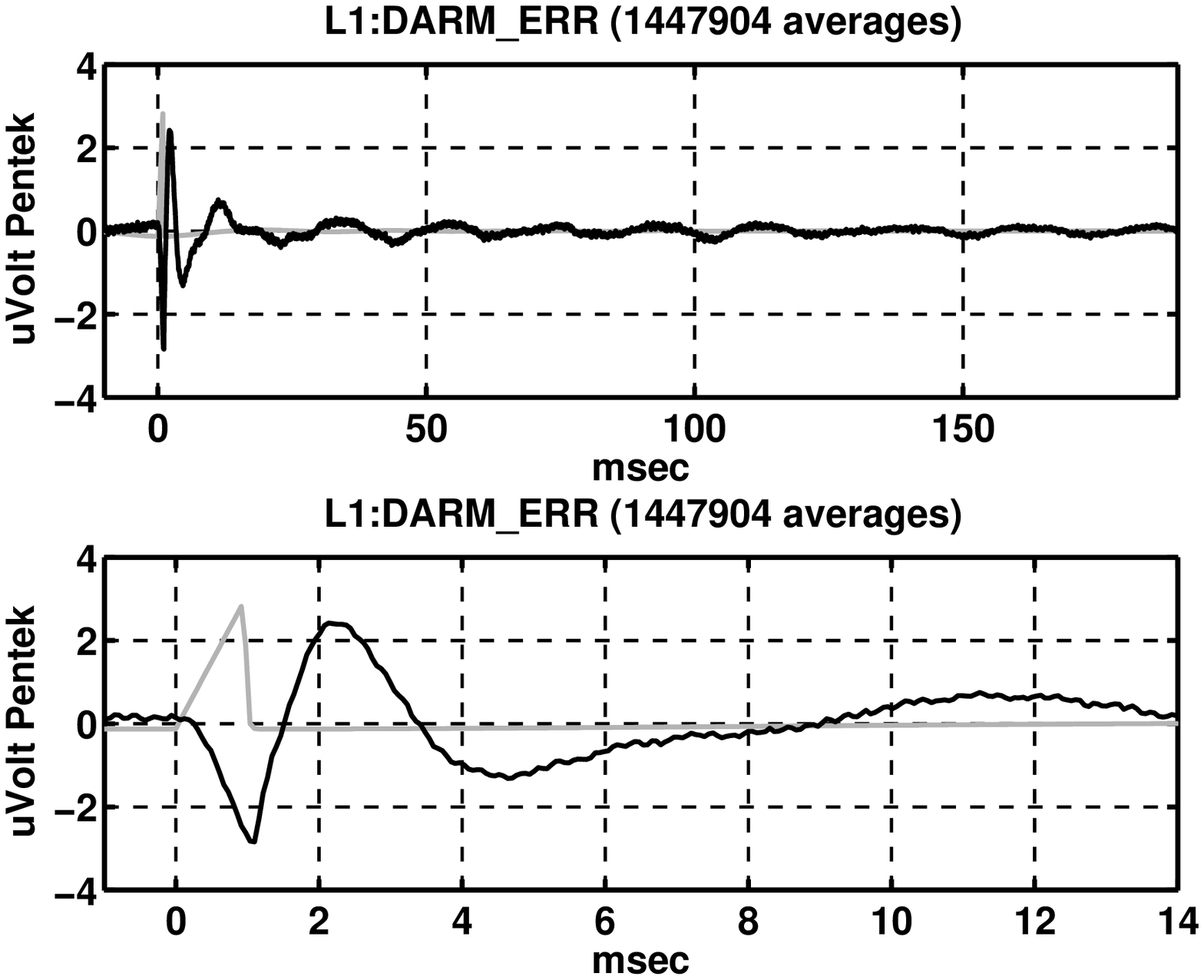}
\end{minipage}
\caption[Periodic timing transient]{{\bf Periodic timing transient} in the gravitational wave channel
         (DARM\_ERR),  calibrated 
         in $\mu{\rm Volt}$ at the ADC (Pentek card)
         for H1 (top two) and L1 (bottom two) shown with a span of 200 msec
         and 14 msec in black. The x-axis is the offset from a full GPS second.
	 About 1.4 million seconds of DARM\_ERR data was averaged to get this trace.
	 Also shown in gray is the GPS\_RAMP
         signal that was used as a timing monitor. It was identified as a cause of the
	 periodic timing transient in DARM\_ERR. The H1 trace shows an additional feature
	 at 6 msec.
	 \label{fig:H1L1TimingTransient}
	 }
\end{figure}

The large spatial separation also reduces environmental correlations
between the two sites. Nevertheless we still found a comb of 1~{\rm Hz} harmonics that was
coherent between H1 and L1. This correlation was found to be
at least in part due to an exactly 1-sec periodic signal in both interferometers
(FIG. \ref{fig:H1L1TimingTransient}), which was caused by cross-talk from the GPS\_RAMP signal.
The GPS\_RAMP signal consists of a 10 msec saw-tooth
signal that starts at every full second, lasts for 1 msec and 
is synchronized with the GPS receivers.
(see FIG. \ref{fig:H1L1TimingTransient}). This ramp was used as an off-line monitor of the
Analog-to-Digital Converter (ADC) card timing and thus was connected to the same ADC card that was used
for the gravitational wave channel, which resulted in a non-zero cross-talk to the gravitational wave channel.

To reduce the contamination from this signal a transient template was subtracted 
in the time domain. This has the advantage that effectively only
a very narrow band ($1/{\rm run time} \approx 1\e{-6}~{\rm Hz}$) is removed around each $1\hz$ harmonic,
while the rest of
the analysis is unaffected. The waveform for subtraction from the raw (uncalibrated)
data was recovered by averaging the data from the whole run in order to produce a typical second.
Additionally, since this typical second only showed significant features in the first 80 msec,
the transient subtraction template was set to zero (with a smooth transition) after 120 msec.
This subtraction was done for only H1 since adding repetitive data to both detectors can introduce
artificial correlation. It eliminated the observed correlation.
However, due to an automatically adjusted gain between the ADC card and the gravitational wave
channel, 
the amplitude of the transient waveform is affected by
a residual systematic error. Its effect on the cross-correlation result was estimated
by comparing maps with the subtraction done on either H1 or L1. The systematic error is mostly
concentrated around the north and south poles, with a maximum of about 50\% of the statistical
error at the south pole. In the declination range of $-75^{\circ}$ to $+75^{\circ}$ the error is less than 10\%
of the statistical error. For upper limit calculations this systematic error is added in quadrature
to the statistical error.
After the S4 run the GPS\_RAMP signal was replaced with a two-tone signal at 900 Hz and 901 Hz.
The beat between the two is now used to monitor the timing.

One post-processing cut was required to deal with detector non-stationarity.
To avoid a bias in the cross-correlation statistics the
segment before and the segment after the one being analyzed
are used for
the power spectral density (PSD) estimate \cite{biasnote}.
Therefore the analysis becomes vulnerable to large, short transients
that happen in one instrument in the middle segment - such transients
cause a significant underestimate of the PSD and thus of the theoretical standard deviation
for this segment. This leads to a contamination of the final estimate.

To eliminate this problem the standard deviation $\sigma$
is estimated for both the middle segment and the two adjacent
segments. The two estimates are then required to agree within 20\%:
\begin{equation}
\label{eq:cut}
\frac{1}{1.2} < \frac{\sigma_{\rm middle}}{\sigma_{\rm adjacent}} < 1.2.
\end{equation}
The analysis is fairly insensitive to the threshold - the only significant contamination comes
from very large outliers that are cut by any reasonable threshold \cite{ballmerthesis}. The chosen threshold of 20\%
eliminates less than 6 \% of the data.

\section{\label{sec:Results}Results from the S4 run}

\subsection{Broadband results}
In this analysis we searched for an $H(f)$ following a power law with two different exponents
$\beta$:
\begin{itemize}
\item {\it $\beta=-3$:
      $H(f)=H_{-3} \left( \frac{100\hz}{f}\right)^{3}$.}\newline This emphasizes low
      frequencies and is useful when interpreting the result in a cosmological framework,
      since it corresponds to a scale-invariant primordial perturbation spectrum, i.e.
      the GW energy per logarithmic frequency interval is constant.
\item {\it $\beta=0$:
      $H(f)=H_{0}$ (constant strain power).}\newline This emphasizes the
      frequencies for which the interferometer strain sensitivity is highest. 
\end{itemize}
The results are reported as point estimate $Y_{\omg}$ and corresponding standard
deviation $\sigma_{\omg}$ for each pixel $\omg$.
The point estimate $Y_{\omg}$ must be interpreted as best fit
amplitude $H_{\beta}$ for the pixel $\omg$ (equation \ref{e:mu2}).

\begin{figure}[!thb]
\centering
\begin{minipage}[t]{2.75in}
\includegraphics[width=1\linewidth]{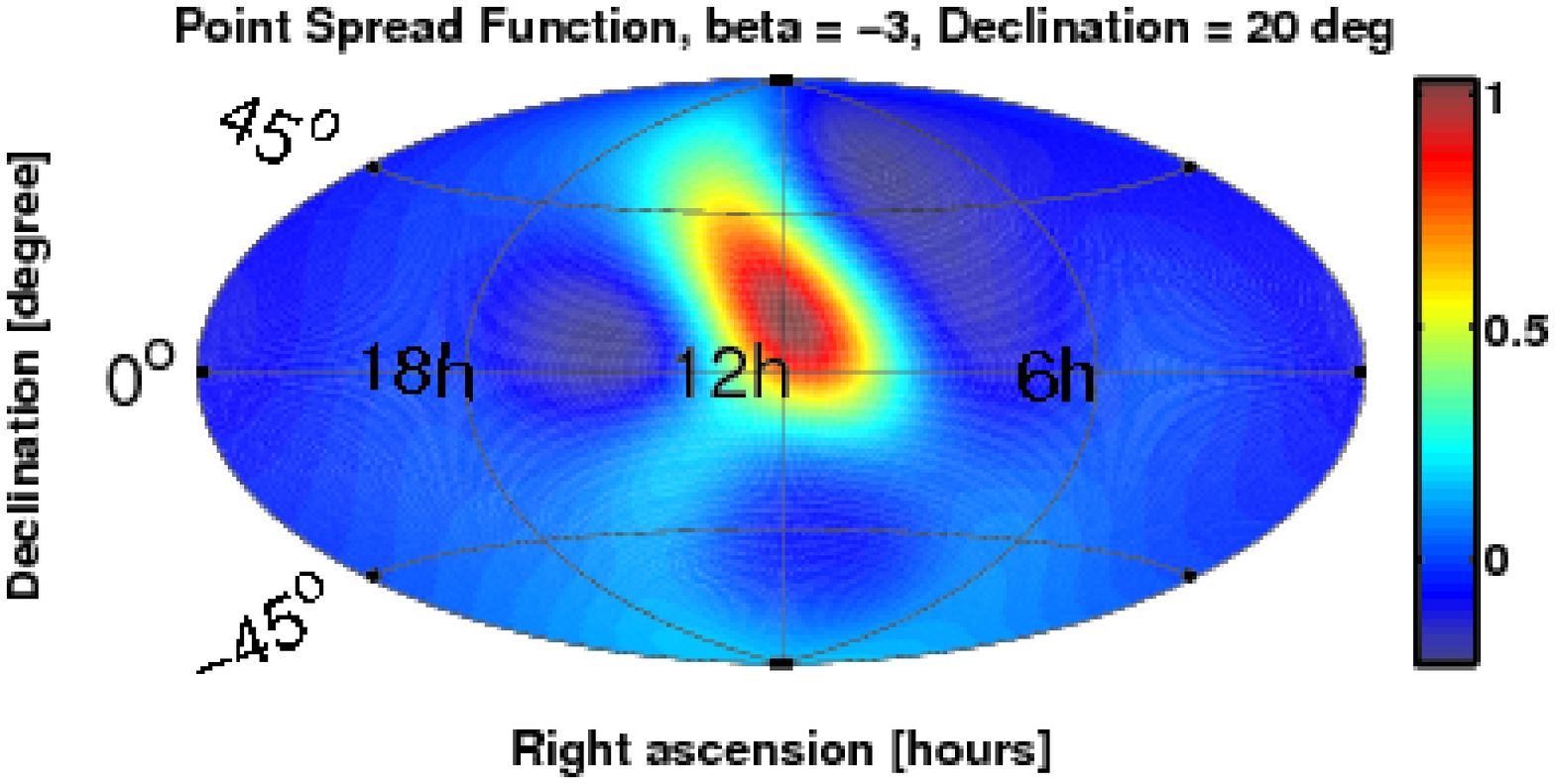}
\includegraphics[width=1\linewidth]{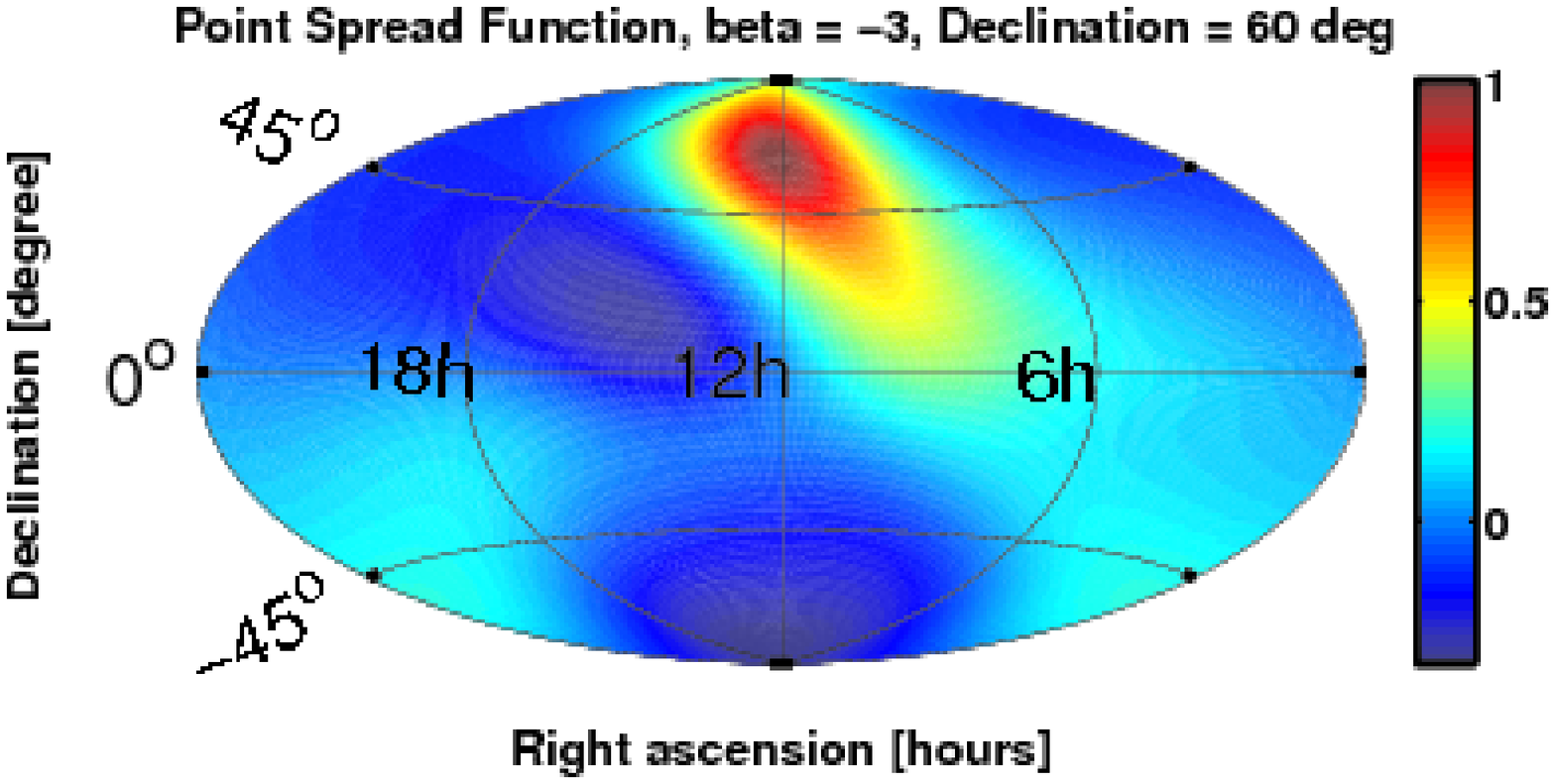}
\includegraphics[width=1\linewidth]{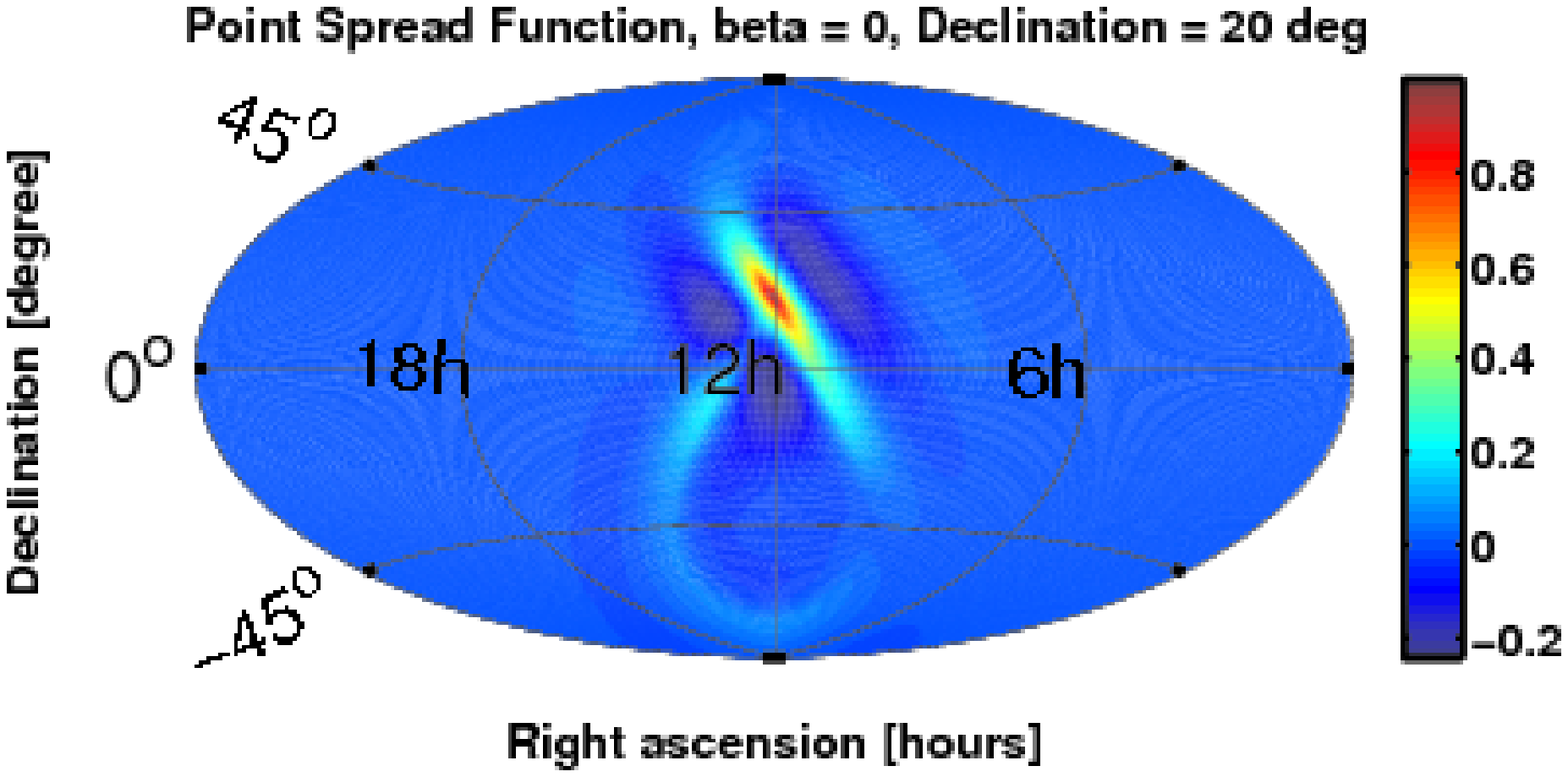}
\includegraphics[width=1\linewidth]{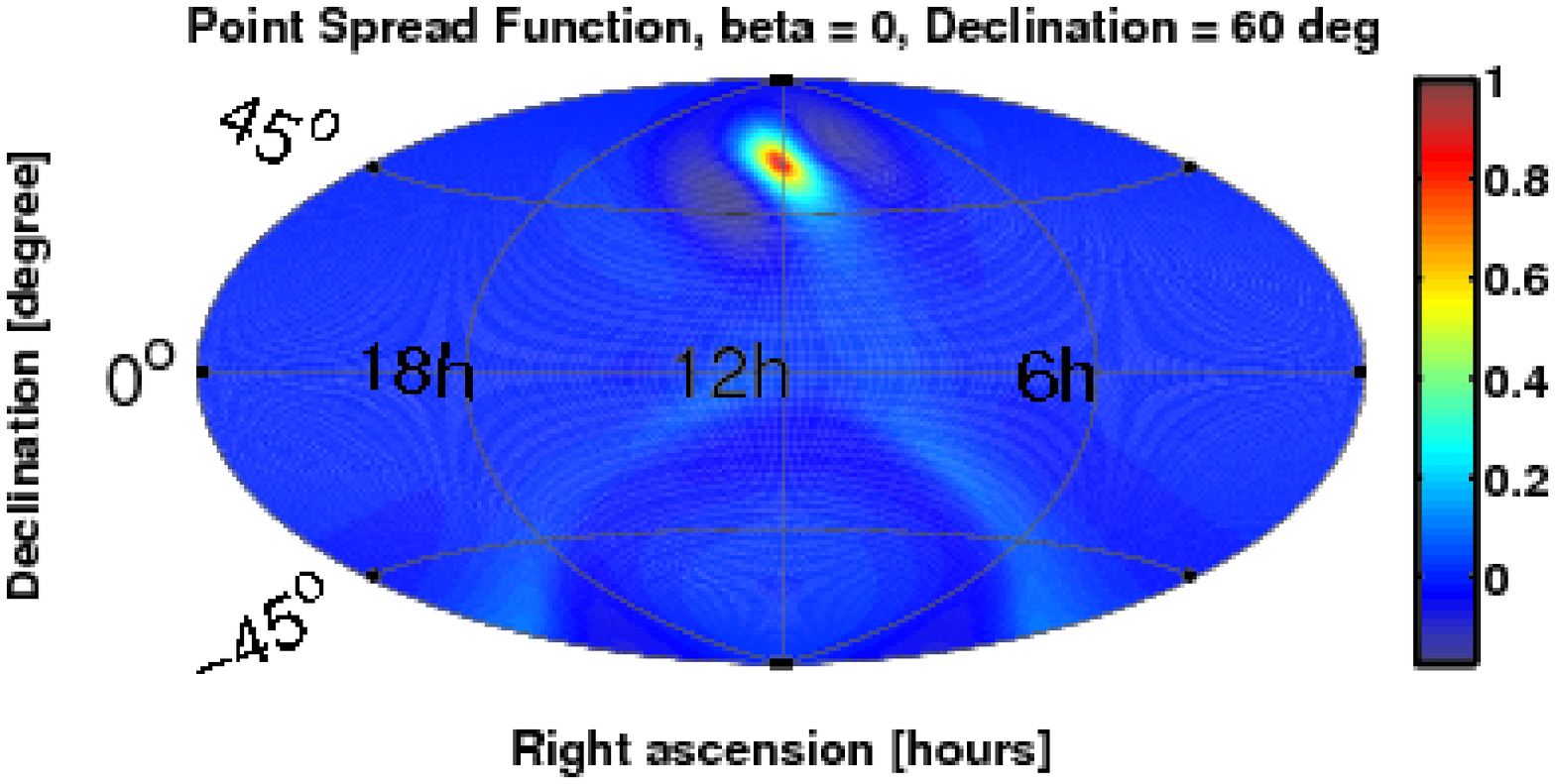}
\end{minipage}
\caption[Antenna lode]{{\bf Point spread function $A(\omg,\omg')$}
         of the radiometer for $\beta=-3$ (top two figures)
         and for $\beta=0$ (bottom two figures). Plotted is the relative
	 expected signal strength assuming a source
	 at right ascension $12~h$ and declinations $20^{\circ}$ and $60^{\circ}$.
	 Uniform day coverage was assumed, so the resulting shapes are independent of right ascension.
	 An Aitoff projection was used to plot the whole sky.
	 \label{fig:PointSpreadFunctionsFunctions}
	 }
\end{figure}

Also we should note that the resulting maps have an intrinsic spatial
correlation, which is described by the point spread function
\begin{equation}\label{e:PointSpreadFunctionsFunction}
    A(\omg,\omg') =
    \frac{\left< Y_{\omg} Y_{\omg'} \right>}{\left< Y_{\omg'} Y_{\omg'} \right>}.
\end{equation}
It describes the spatial correlation in the following sense:
if either $Y_{\omg'}=\bar{Y}$ due to random fluctuations, or
if there is a true source of strength $\bar{Y}$ at $\omg'$,
then the expectation value at $\omg$
is $\left< Y_{\omg} \right> = A(\omg,\omg') \bar{Y}$.
The shape of $A(\omg,\omg')$ depends strongly on the declination.
FIG. \ref{fig:PointSpreadFunctionsFunctions} shows $A(\omg,\omg')$
for different source declinations and both the $\beta=-3$ and $\beta=0$
case, assuming continuous day coverage.

\subsubsection{Scale-invariant case, $\beta=-3$}
\label{subsec:ConstOmegaUL}

\begin{figure}[!thb]
\includegraphics[width=1\linewidth]{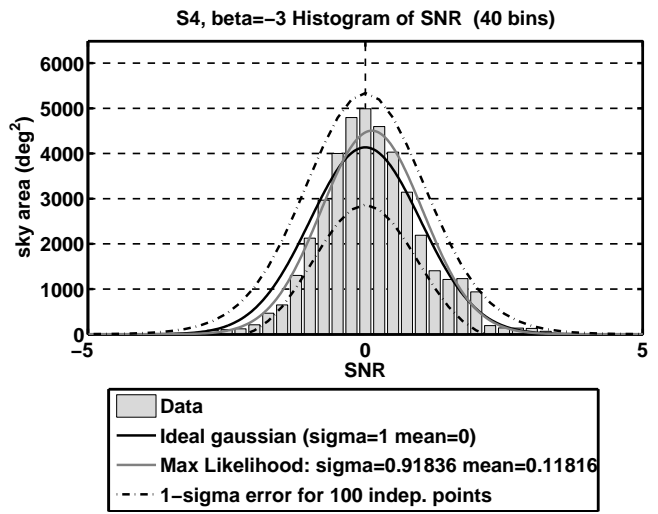}
\caption[S4 Result SNR]{
         {\bf S4 Result:} Histogram of the
         signal-to-noise ratio (SNR) for $\beta=-3$.
      The gray curve is a maximum likelihood Gaussian fit to the data.
      The black solid line is an ideal Gaussian, the two dash-dotted black lines
      indicate the expected one sigma variations around this ideal Gaussian for
      100 independent points ($N_{\rm eff}=100$).
      \label{fig:S4f3SNR}
      }
\end{figure}

A histogram of the
${\rm SNR}=\frac{Y}{\sigma}$ is plotted in
FIG. \ref{fig:S4f3SNR}.
The data points were weighted with the corresponding sky area
in square degrees.
Because neighboring points are correlated, 
the effective number of independent points $N_{\rm eff}$ is reduced.
Therefore the histogram
can exhibit statistical fluctuations that are significantly larger than
those naively expected from simply counting the number of pixels in the map,
while still being 
consistent with (correlated)
Gaussian noise. Indeed the histogram in FIG. \ref{fig:S4f3SNR} features a slight
bump around SNR=2, but is still consistent with $N_{\rm eff}=100$ - the red
dash-dotted lines indicate the one sigma band around the red ideal Gaussian
for $N_{\rm eff}=100$.
Additionally
the SNR distribution also passes a Kolmogorov-Smirnov test for $N_{\rm eff}=100$
at the 90\% significance level.

The number of independent points $N_{\rm eff}$, which in effect describes the diffraction
limit of the LIGO detector pair,
was estimated by 2 heuristic methods:
\begin{itemize}
\item {\it Spherical harmonics decomposition} of the SNR map. The resulting
    power vs $l$ graph shows structure up to roughly $l=9$ and falls off
    steeply above that - the $l=9$ point corresponds to one twentieth
    of the maximal power. The effective number of independent points then
    is $N_{\rm eff}\approx (l+1)^2=100$.
\item {\it FWHM area} of a strong injected source, which is latitude dependent
    but of the order of $800$ square degrees.
    To fill the sky we need about $N_{\rm eff} \approx 50$ of those patches.
    We used the higher estimate $N_{\rm eff}=100$ for this discussion.
\end{itemize}

FIG. \ref{fig:S4f3SNR} suggests that the data is consistent with no signal.
Thus we calculated a Bayesian 90\% upper limit for each sky direction.
The prior was assumed to be flat between zero and an upper cut-off set to
$5\e{-45} {\rm Hz}^{-1}$ at $100\hz$, the approximate limit that can be set
from just operating a single LIGO interferometer at the S4 sensitivity.
Note, however, that this cut-off is so high that the upper limit is completely 
insensitive to it.
Additionally we marginalized over the calibration uncertainty of 8\% for H1 and 5\%
for L1 using a Gaussian probability distribution.
The resulting upper limit map is shown in FIG. \ref{fig:S4f3UL90}.
The upper limits on the strain power spectrum $H(f)$ vary between
$1.2\e{-48} {\rm Hz}^{-1} \left({100~{\rm Hz}}/{f}\right)^3$
and
$1.2\e{-47} {\rm Hz}^{-1} \left({100~{\rm Hz}}/{f}\right)^3$,
depending on the position in the sky. These strain limits correspond
to limits on the gravitational wave
energy flux $F(f)$
varying between
$3.8\e{-6} {\rm erg}~{\rm cm}^{-2}~{\rm Hz}^{-1} \left({100~{\rm Hz}}/{f}\right)$
and
$3.8\e{-5} {\rm erg}~{\rm cm}^{-2}~{\rm Hz}^{-1} \left({100~{\rm Hz}}/{f}\right)$.

\subsubsection{Constant strain power, $\beta=0$}
\label{subsec:ConstStrainPowerUL}

\begin{figure}[!thb]
\includegraphics[width=1\linewidth]{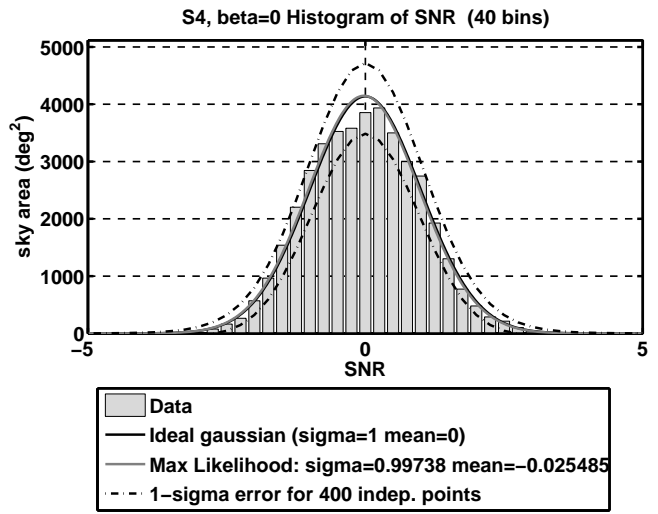}
\caption[S4 Result SNR]{{\bf S4 Result:} Histogram of the
         signal-to-noise ratio (SNR) for $\beta=0$.
	 The gray curve is a maximum likelihood Gaussian fit to the data.
         The black solid line is an ideal Gaussian, the two dash-dotted black lines
         indicate the expected one sigma variations around this ideal Gaussian for
         400 independent points ($N_{\rm eff}=400$).
	 \label{fig:S4flatSNR}
	 }
\end{figure}

Similarly, FIG. \ref{fig:S4flatSNR} shows a histogram of the
${\rm SNR}=\frac{Y}{\sigma}$ for the constant strain power case.
Structure in the spherical harmonics power spectrum goes up
to $l=19$, thus $N_{\rm eff}$ was estimated to be
$N_{\rm eff}\approx (l+1)^2=400$. Alternatively the FWHM area of a strong injection covers
about $100^{{\circ}2}$ which also leads to $N_{\rm eff}\approx 400$. The dash-dotted
red lines in the histogram (FIG. \ref{fig:S4flatSNR}) correspond to the expected $1-\sigma$
deviations from the ideal Gaussian for $N_{\rm eff}= 400$. The histogram is thus
consistent with (correlated) Gaussian noise, indicating that there is no signal present.
The SNR distribution also passes a Kolmogorov-Smirnov test for $N_{\rm eff}=400$
at the 90\% significance level.

Again we calculated a Bayesian 90\% upper limit for each sky direction, including the
marginalization over the calibration uncertainty.
The prior was again assumed to be flat between $0$ and 
an upper cut-off of $5\e{-45} {\rm Hz}^{-1}$ at $100\hz$.
The resulting upper limit map is shown in FIG. \ref{fig:S4flatUL90}.
The upper limits on the strain power spectrum $H(f)$ vary between
$8.5\e{-49} {\rm Hz}^{-1}$
and
$6.1\e{-48} {\rm Hz}^{-1}$
depending on the position in the sky. This corresponds to limits on the gravitational wave
energy flux $F(f)$
varying between
$2.7\e{-6} {\rm erg}~{\rm cm}^{-2}~{\rm Hz}^{-1} \left({f}/{100~{\rm Hz}}\right)^2$
and
$1.9\e{-5} {\rm erg}~{\rm cm}^{-2}~{\rm Hz}^{-1} \left({f}/{100~{\rm Hz}}\right)^2$.

\subsubsection{Interpretation}

\begin{figure}[htbp!]
\begin{center}
  \includegraphics[width=1\linewidth,angle=0]{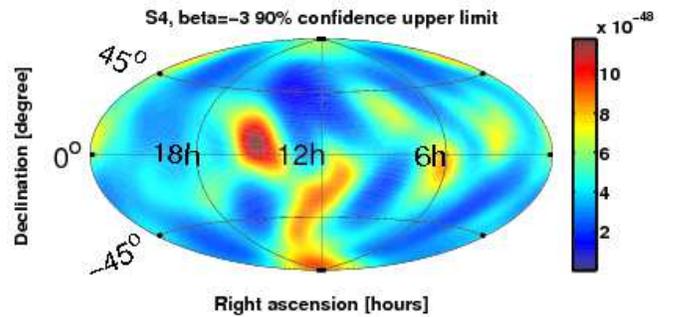}
  \caption[S4 result 90\% confidence level Bayesian upper limit]{
           {\bf S4 Result:} Map of the 90\% confidence level Bayesian upper limit
	   on $H_{\beta}$ for $\beta=-3$. The upper limit varies between
           $1.2\e{-48} {\rm Hz}^{-1} \left({100~{\rm Hz}}/{f}\right)^3$
           and
           $1.2\e{-47} {\rm Hz}^{-1} \left({100~{\rm Hz}}/{f}\right)^3$,
           depending on the position in the sky. All fluctuations are consistent with the expected noise.}
  \label{fig:S4f3UL90}
\end{center}
\end{figure}

\begin{figure}[htbp!]
\begin{center}
  \includegraphics[width=1\linewidth,angle=0]{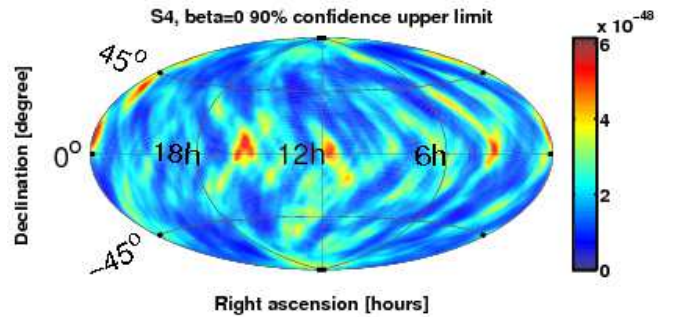}
  \caption[S4 result 90\% confidence level Bayesian upper limit]{
           {\bf S4  Result:} Map of the 90\% confidence level Bayesian upper limit
	   on $H_{\beta}$
           for $\beta=0$.The upper limit varies between
           $8.5\e{-49} {\rm Hz}^{-1}$ and $6.1\e{-48} {\rm Hz}^{-1}$
           depending on the position in the sky.}
  \label{fig:S4flatUL90}
\end{center}
\end{figure}

The maps presented in FIGS. \ref{fig:S4f3UL90} and \ref{fig:S4flatUL90}
represent the first directional upper limits on a stochastic gravitational wave background
ever obtained. They are consistent with no gravitational wave background
being present.
This search is optimized for well localized, broadband sources of gravitational waves.
As such it is best suited for unexpected, poorly modeled sources.

In order to compare the result to what could be expected from known sources we also
search for the gravitational radiation from low-mass X-ray binaries (LMXBs).
They are accretion-driven spinning neutron stars, i.e. narrow-band sources and thus
not ideal for this broadband search. However they have the advantage that we can predict
the gravitational luminosity based on the known X-ray flux.
If gravitational radiation provides the torque balance for LMXBs, then there is a simple
relation between the gravitational luminosity $L_{GW}$ and X-ray luminosity $L_X$ \cite{wagoner}:
\begin{equation}
\label{eq:LMXBbalance}
L_{GW} \approx \frac{f_\text{spin}}{f_\text{Kepler}} L_X.
\end{equation}
Here $f_\text{Kepler}$ is final orbital frequency of the accreting matter, about $2~{\rm kHz}$
for a neutron star, and $f_\text{spin}$ is the spin frequency.

As an example we estimate the gravitational luminosity of all LMXBs within the
Virgo galaxy cluster. Their integrated X-ray luminosity is
about $10^{-9}$ erg/sec/$\text{cm}^2$ (3000 galaxies at 15~Mpc, $10^{40}$~erg/sec/galaxy from LMXBs).
For simplicity we assume that the ensemble  produces a flat strain power spectrum $H(f)$ over a
bandwidth $\Delta f$.
Then the strength of this strain power spectrum is about
\begin{equation}
\label{eq:intf2H_Virgo}
\begin{split}
 H(f) &= \frac{2 G}{\pi c^3} \frac{1}{f_\text{Kepler} f_\text{center} \Delta f} L_{X} \\
 &\approx 10^{-55} \hz^{-1} \left( \frac{100 \hz}{f_\text{center}} \right) \left( \frac{100 \hz}{\Delta f} \right).
\end{split}
\end{equation}
Here $f_\text{center}$ is the typical frequency of the $\Delta f$ wide band of interest.
This is quite a bit weaker than the upper limit set in this paper, which is mostly due to
the fact that the intrinsically narrow-band sources are diluted over a broad frequency band.


\subsection{Limits on isotropic background}

It is possible to recover the  estimate for an isotropic background
as an integral over the map (see \cite{ballmer}).
The corresponding theoretical standard deviation would require a double integral with
essentially the point spread function as integrand. In practice it is simpler to 
calculate this theoretical standard deviation directly by using the overlap reduction function
for an isotropic background (see \cite{allenromano}).
From that the 90\% Bayesian upper limit can be calculated, which is additionally
marginalized over the calibration uncertainty.
In the $\beta=-3$ case the 90\% upper limit we can set on $h_{72}^2 \Omega_{\rm gw}(f)$ is $1.20\e{-4}$.
The dimensionless quantity $\Omega_{\rm gw}(f)$ is the GW energy density per unit logarithmic frequency,
divided by the critical energy density $\rho_c$ to close the universe, and $h_{72}$ is the Hubble constant
in units of $72~{\rm km}~{\rm sec}^{-1}{\rm Mpc}^{-1}$.
Table \ref{t:isotropicS4} summarizes the results for all choices of $\beta$. 
\begin{table*}
\begin{ruledtabular}
\begin{tabular}{lll}
\multicolumn{3}{l}
{{\bf S4 isotropic upper limit}}\\ 
Quantity & $\Omega_{\rm gw}(f) = const$&
  $S_{\rm gw}(f) = const$\\ \hline 
point estimate $Y$		& $1.02\e{-47} \hzff{-3}$& $-7.12\e{-48}{\rm Hz}^{-1}$\\ 
standard deviation $\sigma$	& $6.97\e{-48}\hzff{-3}$	& $~7.22\e{-48}{\rm Hz}^{-1}$\\
\hline
90\% Bayesian UL on $S_{\rm gw}(f)$&
				 $1.99\e{-47}\hzff{-3}$	& $8.49\e{-48}{\rm Hz}^{-1}$\\ 
90\% Bayesian UL on $h_{72}^2 \Omega_{\rm gw}(f)$&
				 $1.20\e{-4}$		& $5.13\e{-5} \ff{3}$\\ 
\end{tabular}
\end{ruledtabular}
\caption[S4 isotropic result]{
          {\bf S4 isotropic result} for the $\Omega_{\rm gw}(f) = const$, ($\beta=-3$) and the 
          $S_{\rm gw}(f) = const$, ($\beta=0$) case. The first two lines show the point estimate and standard
	  deviation that are used to calculate the 90\% Bayesian upper limits. The
	  upper limits are also marginalized over the calibration uncertainty.
	  These results agree
	  with the ones published in \cite{vuk} within the error bar of the measurement.}
\label{t:isotropicS4}
\end{table*}

\subsubsection{Interpretation}

In \cite{vuk} we published an upper limit
of $h_{72}^2\Omega_{GW}<6.5\e{-5}$
on an isotropic gravitational wave background
using S4 data. That analysis is
mathematically identical to inferring the point estimate as an integral
over the map \cite{ballmer}, but the mitigation of the timing transient
and the data quality cuts were sufficiently different to affect the point
estimate.
While both results are consistent within the error bar of the measurement,
this difference results in a slightly higher
upper limit. Both results are significantly better
than the previously published LIGO S3 result. 

\subsection{Narrow-band results targeted on Sco-X1}

As an application we again focus on low-mass X-ray binaries (LMXBs).
The gravitational wave flux from all LMXBs is expected to be dominated by the closest one, Sco-X1,
simply because Sco-X1 dominates the X-ray flux from all LMXBs, and X-ray luminosity $L_X$ is related to
the gravitational luminosity $L_{GW}$ through equation \ref{eq:LMXBbalance}. 
Unfortunately the spin frequency of Sco-X1 is not known.
We thus
want to set an upper limit for each frequency bin on the RMS strain coming from the direction of Sco-X1
(RA: 16h 19m 55.0850s; Dec: $\text{-15}^{\circ}$ 38' 24.9'').

The
binary orbital velocity of Sco-X1 is about $40\pm5~{\rm km/sec}$ (see \cite{vorbit}).
This induces a maximal frequency shift of $\Delta f_{GW} = 2.7\e{-4} \times f_{GW}$.
We chose a bin width of $df=0.25~{\rm Hz}$, which is broader than maximal frequency shift $\Delta f_{GW}$
for all frequencies below $926\hz$ and is the same bin width
that was used for the broadband analysis. Above $926\hz$ the signal is guaranteed to spread over multiple bins.

\begin{figure}[!thb]
\includegraphics[width=1\linewidth]{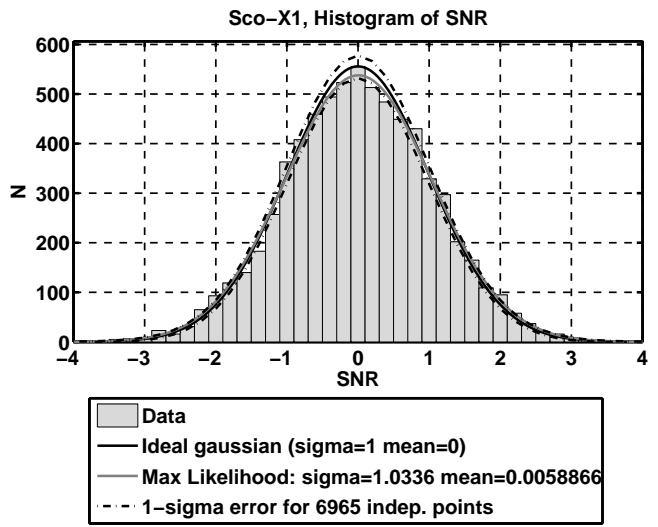}
\caption[S4 Result for Sco-X1, SNR]{{\bf S4 Result for Sco-X1:}
         Histogram of the signal-to-noise ratio calculated for
	 each $0.25~{\rm Hz}$ wide frequency bin.
	 There are no outliers.\label{fig:S4ScoX1UL}
	 }
\end{figure}

To avoid contamination from the hardware-injected pulsars,
the 2 frequency bins closest to a pulsar frequency
were excluded. Multiples of 60 Hz were also excluded.
The lowest frequency bin was at 50 Hz, the highest one at 1799.75 Hz.
FIG. \ref{fig:S4ScoX1UL} shows a histogram of the remaining 6965 $0.25~{\rm Hz}$ wide frequency bins.
It is consistent with a Gaussian distribution (Kolmogorov-Smirnov test with $N=6965$ at the 90\% significance level).

\begin{figure}[!thb]
\includegraphics[width=1\linewidth]{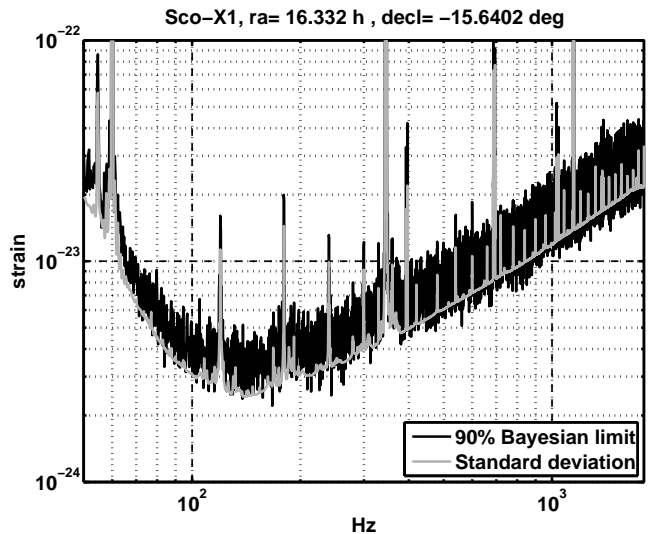}
\caption[S4 Result for Sco-X1, SNR]{{\bf S4 Result for Sco-X1:}
         The 90\% confidence Bayesian upper limit as a function of
	 frequency - marginalized over the calibration uncertainty.
	 The standard deviation (one sigma error bar) is shown in blue.
	 \label{fig:S4ScoX1ULspec}
	 }
\end{figure}

A 90\% Bayesian upper limit for each frequency bin was calculated based on the point estimate and
standard deviation, including a marginalization over the calibration uncertainty.
Figure \ref{fig:S4ScoX1ULspec} is a plot of this 90\%  limit (red trace).
Above about 200 Hz (shot noise regime above the cavity pole)
the typical upper limit rises linearly with frequency and is given by
\begin{equation}
\label{eq:SCOX1UL}
h_{\rm RMS}^{(90\%)} \approx 3.4\e{-24} \left( \frac{f}{200~{\rm Hz}} \right)\,\,\,\,f \gsim 200~{\rm Hz}.
\end{equation}
The standard deviation is also shown in blue.

\subsubsection{Interpretation}

In principle,
the radiometer analysis is not an optimal method to search for
a presumably periodic source like Sco-X1.
Nevertheless it can set a competitive upper limit with a minimal set of assumptions on the source
and significantly less computational resources.
Indeed LIGO published a 95\% upper limit on gravitational radiation amplitude from Sco-X1 of
$1.7\e{-22}$ to $1.3\e{-21}$ across the $464-484\hz$ and $604-624\hz$ frequency
bands \cite{s2scox1}, using data from S2,
which had a noise amplitude about $4.5$ times higher around $500\hz$ in each instrument.
The analysis was computationally limited to using $6$ hours of data and two $20\hz$ frequency bands.
However the strain amplitude
sensitivity scales as $T^{-1/4}$ \cite{ballmer}, while a coherent method scales as $T^{-1/2}$. 

The upper limit (eq. \ref{eq:SCOX1UL}) can directly be compared to the expected strain based on the
X-ray luminosity:
\begin{equation}
\label{eq:SCOX1aboveexp}
\frac{h_{\rm RMS}^{(90\%)}}{h_{\rm RMS}^{L_X}} \approx
 100 \left( \frac{f}{200~{\rm Hz}} \right)^{\frac{3}{2}}\,\,\,\,f \gsim 200~{\rm Hz}.
\end{equation}
Here $f$ is the gravitational wave frequency, i.e. twice the (unknown) spin frequency of Sco-X1.
This is close enough that,
if the model described in \cite{wagoner}, and thus equation \ref{eq:LMXBbalance} are indeed correct, 
Sco-X1 ought to be detectable with this method and the next generation of gravitational
wave detectors operated in a narrow-band configuration (AdvLIGO \cite{AdvLIGO}).
For a discussion of the expected signal from Sco-X1 see also \cite{s2scox1}.

\section{\label{sec:Conclusion}Conclusion}

We produced the first upper limit maps for a stochastic gravitational wave background
by applying a method
that is described in \cite{ballmer}
to the data from the LIGO S4 science run. No 
signal
was seen and
upper limits were set for two different choices for the strain power spectrum $H(f)$.
In the case of $H(f)\propto f^{-3}$ the upper limits for a point source vary between
$1.2\e{-48} {\rm Hz}^{-1} \left({100~{\rm Hz}}/{f}\right)^3$
and 
$1.2\e{-47} {\rm Hz}^{-1} \left({100~{\rm Hz}}/{f}\right)^3$,
depending on the position in the sky (see FIG. \ref{fig:S4f3UL90}).
Similarly, in the case of constant $H(f)$ the upper limits vary between
$8.5\e{-49} {\rm Hz}^{-1}$
and
$6.1\e{-48} {\rm Hz}^{-1}$ (see FIG. \ref{fig:S4flatUL90}).
As a side product limits on an isotropic background of gravitational waves were also obtained, see
TABLE \ref{t:isotropicS4}.

In an additional application, narrow-band upper limits were set
on the gravitational radiation coming from the closest
low-mass X-ray binary, Sco-X1 (see FIG. \ref{fig:S4ScoX1ULspec}). In the shot noise limited
frequency band (above about $200~{\rm Hz}$) the limits on the strain in each $0.25\hz$ wide frequency bin
follow roughly
\begin{equation}
h_{\rm RMS}^{(90\%)} \approx 3.4\e{-24} \left( \frac{f}{200~{\rm Hz}} \right)\,\,\,\,f \gsim 200~{\rm Hz},
\end{equation}
where $f$ is the gravitational wave frequency (twice the spin frequency).

\begin{acknowledgments}
The work described in this paper was part of the doctoral thesis of
S.~W.~Ballmer at the Massachusetts Institute of Technology \cite{ballmerthesis}.
Furthermore
the authors gratefully acknowledge the support of the United States National 
Science Foundation for the construction and operation of the LIGO Laboratory 
and the Particle Physics and Astronomy Research Council of the United Kingdom, 
the Max-Planck-Society and the State of Niedersachsen/Germany for support of 
the construction and operation of the GEO600 detector. The authors also 
gratefully acknowledge the support of the research by these agencies and by the 
Australian Research Council, the Natural Sciences and Engineering Research 
Council of Canada, the Council of Scientific and Industrial Research of India, 
the Department of Science and Technology of India, the Spanish Ministerio de 
Educacion y Ciencia, The National Aeronautics and Space Administration, 
the John Simon Guggenheim Foundation, the Alexander von Humboldt Foundation, 
the Leverhulme Trust, the David and Lucile Packard Foundation, 
the Research Corporation, and the Alfred P. Sloan Foundation.
This paper has been assigned the LIGO document number LIGO-P060029-00-Z.

\end{acknowledgments}





\end{document}